\documentclass[12pt,showpacs,preprintnumbers,amsmath,amssymb]{revtex4}

\usepackage{graphicx}
\usepackage{dcolumn}
\usepackage{bm}
\usepackage{epsfig}

\begin{document}

\title{Superconducting quantum interference phenomenon in Bi$_2$Sr$_2$CaCu$_2$O$_{8+\delta}$ single crystals}

\author{M. Sandberg$^1$}
\author{V. M. Krasnov$^{2}$}
\email{vladimir.krasnov@physto.se}
\address{$^1$ Department of Microtechnology and Nanoscience, Chalmers
University of Technology, SE-41296 G\"oteborg, Sweden\\
$^2$ Department of Physics, Stockholm University, Albanova
University Center, SE-10691 Stockholm, Sweden}

\date{\today}

\begin{abstract}
The operational dc-SQUID based on intrinsic Josephson junctions in
Bi$_2$Sr$_2$CaCu$2$O$_{8+\delta}$ high-$T_c$ superconductor is
fabricated and studied. The novel in-plane loop layout and the
developed in-situ endpoint detection method allowed an accurate
control of the number of junctions in the SQUID. A clear periodic
modulation of the superconducting current as a function of
magnetic flux through the SQUID loop is observed. This is an
unambiguous evidence for the quantum interference phenomenon in
Bi$_2$Sr$_2$CaCu$2$O$_{8+\delta}$ single crystals.
\end{abstract}

\pacs{85.25.Dq, 74.72.Hs, 74.50.+r}
\maketitle

Atomic scale "intrinsic" Josephson junctions (IJJ's) are naturally
formed in layered high-$T_c$ superconductors such as the
Bi$_2$Sr$_2$CaCu$_2$O$_{8+\delta}$ (Bi-2212)
\cite{Muller,Walenh,Schlen,Fiske,Wang,Latysh,Krasnov_TH,Fluctuation}.
IJJ's can be employed as building blocks for various
cryoelectronic devices \cite{Wang,Latysh}. Record-large $I_c R_n$
values $\sim$ 10-15 mV make them particularly attractive for high
frequency applications \cite{Walenh,Wang}. IJJ's are perfect for
3D-integration of many junctions. Operating 3D quantum devices
containing 11000 IJJ's were already demonstrated \cite{Wang}.
Moreover, properties of IJJ's can be controlled in a wide range by
changing O-doping \cite{Doping}, temperature, magnetic field and
intercalation \cite{Krasnov_TH}. This is important for device
applications and may allow tunability similar to band engineering
in semiconducting quantum devices \cite{QCL}.

The Superconducting Quantum Interference Device (SQUID) is one of
the most important and widely used cryoelectronic devices
\cite{Koelle}, both due to it's extreme sensitivity used in
various types of sensors, and as a basis element for digital
Josephson electronics \cite{RSFQ}.

Operation of a dc-SQUID, based on IJJ's, was analyzed in
Ref.\cite{SquidIJJ}. Since the interlayer spacing $\simeq 15 \AA$
in Bi-2212 is very small, such an "intrinsic" SQUID will contain
several stacked IJJ's in each arm. It is known that operation of a
multi-junction SQUID is complicated due to existence of metastable
states and a small current-flux modulation \cite{Lewand}.
Basically, the same is true for a stacked SQUID, even though some
improvement can be achieved \cite{SquidIJJ} in case of
phase-locking of IJJ's in the stack \cite{Mros}. Anyway, the
current-flux modulation of the stacked SQUID decreases roughly
inversely proportional to the number of junctions \cite{SquidIJJ}.
Therefore, SQUID's should have as few and as identical IJJ's as
possible to have the best performance. In the previous attempt, a
3D Focused Ion Beam (FIB) sculpturing was used to fabricate the
intrinsic SQUID with the out-of-plane (in the $bc-$plane) loop.
The SQUID contained a large amount of junctions and no
interference pattern could be observed \cite{Kim}.

Here we report on successful fabrication of operational Bi-2212
intrinsic dc-SQUID with the novel in-plane loop layout, made by
double-side patterning and FIB milling. The progress was achieved
by developing an accurate end-point detection technique, which
allowed fabrication of the SQUID with only few IJJ's. The
fabricated device shows periodic modulation of the superconducting
current as a function of applied magnetic field, which is a clear
evidence for the quantum interference phenomenon in Bi-2212 single
crystals \cite{Irie}.

Fig. 1a) shows a layout of the intrinsic dc-SQUID with the
in-plane loop \cite{SquidIJJ}. Here the SQUID loop is placed in
the $ab$-plane and stacks of IJJ's are formed by etching trenches
from top and bottom sides of the crystal.

Figs. 1 b,c) show a secondary electron- and a back-illumination
optical images, respectively, of the actual intrinsic dc-SQUID
studied in this work. The fabrication procedure required a double
side fabrication technique \cite{Wang} and consisted of five major
steps:

i) A Bi-2212 crystal was glued to a sapphire substrate using hard
baked photoresist, the top part was cleaved-off using adhesive
tape and 30 nm of Au was deposited to prevent surface
deterioration.

ii) A large mesa structure $120 \times 120 \mu m^2$ with a hight
of $\sim 0.5 \mu m$ was etched on top of the crystal by wet
chemical etching. A bottom trench, line - (C) in Fig. 1 b), was
etched using a combination of argon ion milling and wet chemical
etching in a saturated EDTA solution.

iii) The crystal was flipped and glued to a new substrate with the
mesa facing down.

iv) The crystal was cleaved-off, leaving only the large mesa on
the substrate. A metallization gold layer was deposited. The SQUID
loop, electrodes and contact pads were formed by photolithography
and argon ion milling. Three SQUID's were made on each chip.

v) The sample was transferred to a standard FIB. The loop and the
top trench, line - (B) in Fig. 1 b), were made by FIB. In
addition, FIB-cuts were made to separate SQUID's made on the same
crystal, see vertical bright lines in Fig. 1 c). Stacks of IJJ's,
restricted by the bottom and the top trenches, were formed in each
arm of the SQUID loop, see areas marked (IJJ's) in Fig. 1 b) and
the layout in Fig. 1 a).

\begin{figure}
\noindent
\begin{minipage}[c]{0.5\textwidth}
\includegraphics[width=8cm]{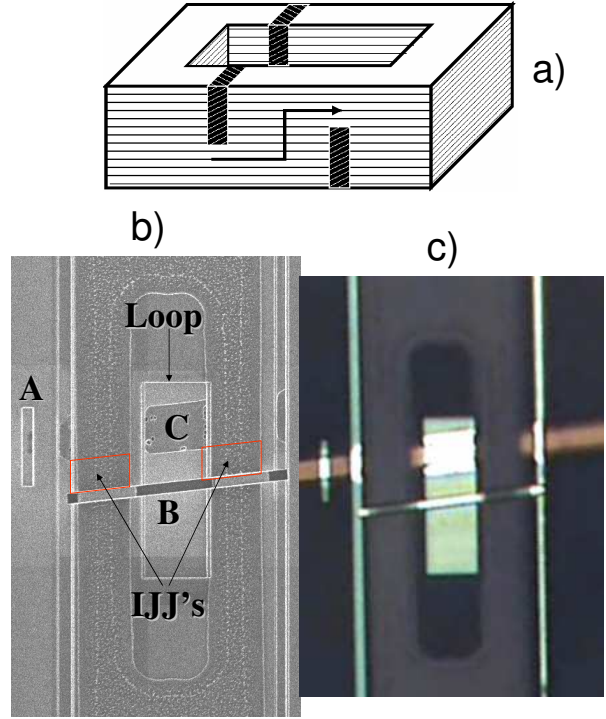}
\end{minipage}
\caption{\label{fig:FIB} a) The layout of intrinsic dc-SQUID with
in-plane loop. The current is forced to flow in the $c-$ axis
direction by trenches from top and bottom sides of SQUID arms. b)
SEM image of the SQUID: A - the small window used for endpoint
detection. B- the top trench, C- the bottom trench seen through
the loop hole. c) back-illumination optical image of the same
SQUID. Etched parts are seen as bright areas. The parameters of
the SQUID are: the loop size $20 \times 7 \mu m^2$, the IJJ stack
sizes in both arms $\sim 5 \times 3 \mu m^2$.}
\end{figure}

The main challenge for fabrication of the SQUID is an accurate
control of the number of IJJ, $N$, in each arm of the SQUID. $N$
is proportional to the overlap between top and bottom trenches. To
minimize $N$ etching of the top trench should be stopped as soon
as it reaches the level of the bottom trench. To do this we
developed a simple endpoint detection method: a test line with the
same width as the top trench was milled simultaneously with the
top trench in the area above the bottom trench, outside the SQUID,
see the line - A in Fig. 1 b). When the test line was etched down
to the bottom trench, secondary electron emission, monitored
during FIB-milling, was reduced and etching was terminated. The
contrast between etched (darker) and unetched areas within the
test line is clearly seen in Fig. 1 b). This simple endpoint
detection method provided an accuracy of $\sim$ few IJJ's.

\begin{figure}
\noindent
\begin{minipage}{0.48\textwidth}
\epsfxsize=.8 \hsize \centerline{ \epsfbox{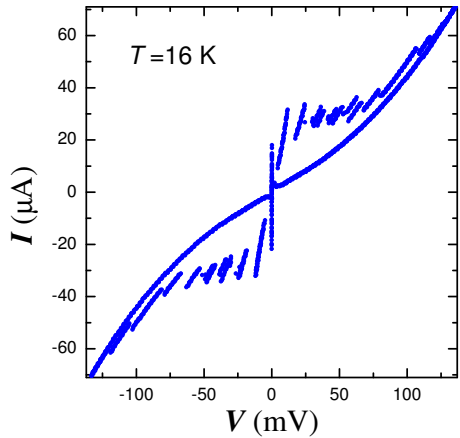} }
\caption{The IVC of the dc-SQUID, shown in Fig.1, at $T \simeq 16
K$. There are six plus one regular quasiparticle branches and a
number of tiny ghost sub-branches caused by parallel connection of
the two slightly different stacks of IJJs. The number of IJJ's in
each stack $N=6+1$.}
\end{minipage}
\end{figure}

Fig. 2 shows the Current-Voltage characteristics (IVC) at $T
\simeq 16 K$ for the same SQUID. The IVC was measured in the
four-probe/superconducting two-probe configuration, avoiding the
contact resistance. A multi-branch structure due to one-by-one
switching of IJJ's from the superconducting into the resistive
state is seen. Counting the number of branches we conclude that
the SQUID contains $N=6$ stacked IJJ's with the critical current
$I_c \simeq 15-20 \mu A$ and one larger junction with $I_c \simeq
25-30 \mu A$, which, however, remained in the superconducting
state during experiments discussed below.

In Fig. 2 we can also see tiny "ghost" sub-branches, which
sometimes are also seen in single stack mesas and are due to
non-uniformity of junctions \cite{Ghost}. However, the
sub-branches in SQUID's are much more pronounced than in mesas.
This is probably due to parallel connection of stacks with
slightly different areas and critical currents. Each time an extra
IJJ switches into the resistive state in one of the stacks, the
current distribution between the arms of the SQUID has to be
re-adjusted to maintain the same voltage across both stacks.

\begin{figure}
\begin{minipage}[c]{0.5\textwidth}
\centering  \includegraphics[width=7.5cm]{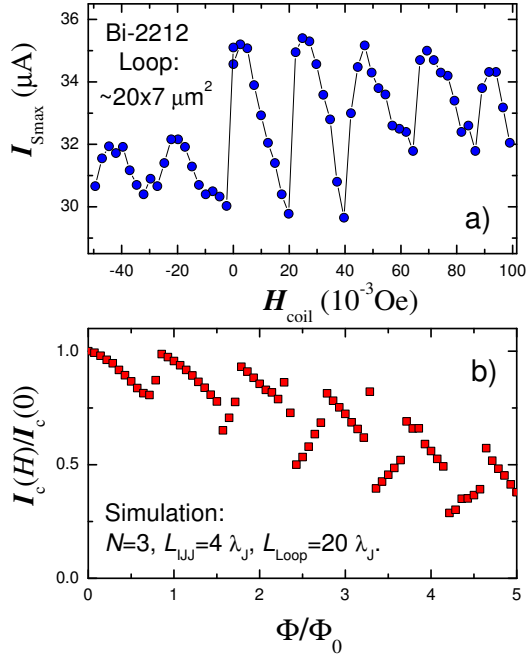}
\end{minipage}
\caption{a) Modulation of the most probable switching current as a
function of applied magnetic field, for the same SQUID as in
Figs.1, 2 at $T \simeq 4.2 K$. Periodic modulation clearly
indicates the quantum interference phenomenon.  b) Numerically
simulated current-flux modulation for a dc-SQUID with $N=3$
identical stacked IJJ's in each arm. A characteristic saw-tooth
like modulation is seen (data from Ref. \cite{SquidIJJ}). }
\label{fig:Mod}
\end{figure}

Fig. 3 a) shows the measured dependence of the most probable
switching current, $I_{Smax}$, on magnetic field, applied
perpendicular to the SQUID loop at $T\simeq 4.2 K$, for the same
device. To obtain $I_{Smax}$, 30720 switching events from the
superconducting to the resistive state were measured using
sample-and-hold technique
\cite{Fluctuation}. This was done to reduces ambiguity caused by
thermal fluctuations, which were significant in this device due to
a small value of $I_c$.

A clear periodic modulation of the switching current vs. the
applied magnetic field is seen in Fig. 3. The period is several
times smaller than the flux quantum divided by the loop area. This
is caused by flux focusing due to a specific geometry of the
device: the SQUID is surrounded by Bi-2212 crystal, seen as dark
areas in Fig. 1 c). Therefore, magnetic field can penetrate only
through the cuts in the crystal, seen as bright vertical lines in
Fig. 1 c), resulting in strong flux focusing effect near the edges
of the SQUID loop. The maximum amplitude of modulation $\sim 15
\%$ is consistent with a rough estimation $\Delta I_c/I_c \sim
1/N$ for $N = 6$ IJJ's in each stack. Moreover, the characteristic
saw-tooth like shape of the modulation is consistent with the
numerical simulations for a stacked SQUID \cite{SquidIJJ}, shown
in Fig. 3 b).

The in-plane SQUID loop layout, employed here has several
important advantages in comparison to the out-of-plane loop layout
used in the previous work \cite{Kim}:

(i) First, slow ion beam (not necessarily focused) etching in the
$c-$axis direction, in combination with the developed endpoint
detection, allows a much better control of the number of IJJ's
(fabrication of a single IJJ is feasible \cite{SingleIJJ}),
compared to FIB etching in the $ab$-plane direction \cite{Kim}.
The number of IJJ's is the most important parameter of the
intrinsic SQUID, since modulation of $I_c$ decreases roughly as
$1/N$ \cite{SquidIJJ}. In our devices no modulation could be seen
for $N > 20$, as it probably became less than thermal fluctuations
and noise in the system.

(ii) Second, the in-plane loop is not limited by the small
thickness of Bi-2212 crystals and, therefore, can be made almost
arbitrary large. The loop inductance can easily be made much
larger than the inductance of IJJ's, improving SQUID performance
\cite{Koelle}. For example, the inductance of the loop for our
SQUID is more than two orders of magnitude larger than for the
out-of-plane SQUID studied in Ref.\cite{Kim}.

In conclusion, we fabricated and studied operational intrinsic
dc-SQUID made from a Bi-2212 single crystal. The observed periodic
modulation of the switching current vs. the magnetic flux through
the SQUID loop is a clear evidence for the quantum interference of
superconducting wave functions in Bi-2212 single crystals. In
total more than ten devices were studied. It was observed that the
amplitude of modulation decreases rapidly with the number of IJJ's
in the SQUID. Therefore, to improve the performance of the
intrinsic SQUID the number of IJJ's should be further reduced.
This can be done using the endpoint detection method developed
here, possibly in combination with the slow Ar-ion milling
technique, using which several groups have demonstrated a
possibility to fabricate single IJJ's \cite{SingleIJJ}. This may
open a possibility for building 3D-integrated electronic circuits
based on intrinsic Josephson junctions.

\begin{acknowledgments}
The work was supported by the Swedish Research Council, grant Nr:
621-2001-3236.
\end{acknowledgments}


\end{document}